# Political Power and Mortality: Heterogeneous Effects of the U.S. Voting Rights Act

December 4, 2025

Atheendar S. Venkataramani (University of Pennsylvania and NBER)
Rourke O'Brien (Yale University)
Elizabeth F. Bair (University of Pennsylvania)
Christopher Lowenstein (University of Missouri – Columbia)

**Abstract:** We study the health consequences of the 1975 extension of the Voting Rights Act, which eliminated key barriers to political participation for previously disenfranchised non-white populations. The reform produced large declines in under-five mortality but sharply divergent effects in other age groups: most non-white cohorts experienced relative mortality reductions, whereas mortality rose among whites and among older non-white men. These patterns are not explained by shifts in population composition or by improvements in material conditions, which accrued to all groups. Instead, the evidence points to changes in perceived status—and the retaliatory behaviors they elicited—as central mechanisms behind the heterogeneous mortality responses.

**Acknowledgements**: We thank Diane Alexander, Mary Bassett, Lisa Berkman, Seth Binder, Susan Busch, Paula Chatterjee, Xi Chen, Damian Clarke, Norma Coe, Jessica Cohen, John Graves, Kevin Griffith, Scott Halpern, Nancy Krieger, L.J. Ristovska, Christina Roberto, Eric Roberts, Krista Ruffini, Mark Schlesinger, Harsha Thirumurthy, Kevin Volpp, Jacob Wallace, Emma Zang, Davene Wright, and seminar participants at Harvard University, Yale University, Vanderbilt University, the University of Pennsylvania, and Penn Behavioral Economics and Health Workshop for helpful comments and suggestions. Venkataramani: atheenv@pennmedicine.upenn.edu; O'Brien: Rourke.obrien@yale.edu; Bair: efbair@pennmedicine.upenn.edu; Lowenstein: clowenstein@missouri.edu.

Policies that extend political power to marginalized groups can have positive effects on their well-being. Several studies have documented meaningful health improvements resulting from policies that expand access to voting or boost representation in government among women and minority groups (Miller 2008; Bhalotra et al. 2023; Lynch 2023). Improvements in health may accrue through several mechanisms, including increased investments in salutary public goods and services, increased public sector accountability, and raising aspirations among the newly enfranchised (Chattopadhyay and Duflo 2004; Beaman et al. 2012).

However, theory suggests that extending political power need not always have positive effects. Policies that do so may invite retaliation against marginalized groups they seek to enfranchise (Acemoglu and Robinson 2008; Phelan et al. 2010; Metzl 2019; Darity Jr 2022), limiting or even reversing any benefits on health or well-being that would otherwise accrue. In addition, relative declines in power experienced by the non-targeted dominant or majority group may worsen their health and well-being. This could occur if extensions of power to marginalized groups lead to redistribution of material resources away from majority groups. Negative effects for non-targeted groups may also materialize through psychological mechanisms: a growing literature has emphasized how intergroup or status threat – the perception that one's social status is at risk or diminished as a result of another group's gain -- may increase stress and directly harm health, just as it may foment retaliation against marginalized groups (Blumer 1958; Blalock 1967; Siddiqi et al. 2019; Efird et al. 2023).

We examine these possibilities by studying the impacts of the 1975 Extension of the U.S. Voting Rights Act (VRA) on mortality. The VRA was first enacted in 1965, expanded access to voting among non-white individuals in 7 states in the Deep South. We focus on the 1975 extension of the VRA, which brought into coverage Texas, Arizona, Alaska and counties in 6 other states, and argue this extension may have led to direct, measurable, and heterogeneous impacts on population health outcomes. The 1975 extension is well suited to address our research question for several reasons. The extension came as a surprise to newly covered jurisdictions (Ang 2019) and, unlike the original VRA, took place outside the wave of major Civil Rights Era and Great Society social policies, many of which have been shown to have health impacts of their own (Bambra et al. 2025). In addition, the salience of the 1975 extension to policymakers and the public was heightened by the impacts of the original VRA – which were well known by the time of the extension (U.S. Commission on Civil Rights 1975) – and its passage was met with immediate backlash from majority groups (Ang 2019; Bernini et al. 2024). Together, these elements allow us to isolate the causal effects of the VRA on health outcomes and make it plausible that these effects varied across racial groups.

Using the universe of U.S. Vital Statistics death certificate data for the period 1970-1988, our empirical approach examines changes in county-level mortality by race and age group before and after 1976 – the first full year in which the VRA extension was in effect -- in newly covered counties versus counties that were not subject to the original or extended VRA. We adopt a synthetic difference-in-differences (SDID) design, which relies on a weaker parallel trends assumption, limits researcher degrees of freedom, and improves efficiency relative to a standard two-way fixed effects (TWFE) approach (Arkhangelsky et al. 2021). We estimate both average and dynamic effects of the VRA, using the latter to help adjudicate underlying mechanisms.

We find substantial heterogeneity in the impacts of the VRA extension on mortality by racial group and age. We first show that mortality gradually declined for children under age 5 for all races in counties covered by the VRA compared to uncovered counties. We also demonstrate relative decreases in mortality for non-white children aged 5-19 and adults aged 20-49. However -- in stark contrast -- we find relative *increases* in mortality for white children ages 5-19 and adults in counties covered by the VRA versus those not covered. These increases are immediate, as we estimate effects that are statistically distinguishable from zero for older white children and adults in the first full year the VRA went into effect. Across the entire post-VRA period, we estimate mortality increases of 5% and 2% for white children ages 5-19 and white adults ages 20-49, respectively, though these overall effects are imprecisely estimated. Estimates for white adults over 50 years of age – implying an average 3% relative increase in mortality in the post-VRA period – are statistically significant. These core findings are overall robust to specification and the inclusion of a range of additional policy covariates.

Disaggregating by cause of death, we show that mortality impacts are primarily driven by changes in cardiovascular disease, strokes, and, to a lesser extent, accidental deaths. We also find evidence of heterogeneity by gender, with the relative magnitude of impacts generally larger for white boys and younger men. For non-white children and adults, the impacts are generally larger in magnitude relative to mean mortality rates for girls and women. For non-white men over age 50, we find evidence of a relative increase in mortality in VRA covered areas, which is mirrored by a decrease among non-white women in the same age group.[1]

We then investigate a range of potential explanations for the striking heterogeneity in health impacts we find. We show that population selection due to selective migration or fertility is unlikely to explain our findings. In fact, shifts in population composition appear to run counter to the estimated health effects of the VRA for most age groups. We next examine changes in material resources and find that these improved for all racial groups. Specifically, we find that counties with both above and below median shares of non-white individuals experienced similar increases in per capita income after the VRA was implemented. Analysis of survey data from the period demonstrates increases in household income in both groups, though these were larger for non-white households in both absolute and relative terms. While we also find evidence of increases in expenditures per capita that are larger in areas with greater non-white populations, consistent with prior work on the VRA by Cascio and Washington (2014), estimates for areas with lower non-white populations are still positive (if imprecisely estimated). Together, these findings for material resources may help explain the declines in mortality experienced by children under 5 for both racial groups, but not the immediate relative increases in mortality experienced by white older children and adults in VRA covered areas or the increases experienced by non-white men over age 50. We further rule out crowd-out from utilization of scarce health care resources by non-white individuals as a mechanism for the mortality increases among whites.

Having found no clear empirical support for these alternate causes, we next consider possibility that the relative mortality increases we find after the VRA for white older children and adults can be explained by psychological mechanisms, namely increased perceptions of threats to status for white individuals and resulting retaliation against non-white men. A growing body of laboratory and real-world evidence in psychology and sociology suggests that decreases in perceived individual or racial group position may affect physiological stress responses that in turn can harm health (Demakakos et al. 2008; Malat et al. 2018; Siddiqi et al. 2019; Rambotti 2022; Efird et al. 2023). Our analysis of the dynamic effects of the VRA supports this possibility:

---

[1] We show that these increases are unlikely to be driven by knowledge of the Tuskegee Study of Untreated Syphilis, which was revealed 3 years prior to the VRA and has been linked to reductions in life expectancy among older Black men (Alsan and Wanamaker 2018).

the immediate relative increases in mortality are more consistent with psychological mechanisms than shifts in material circumstances, which tend to evolve more slowly (Boen et al. 2025; Venkataramani et al. 2025), particularly when it involves the provision of public goods, which are subject to budgeting and electoral cycles.

We provide several additional pieces of evidence consistent with the idea that extension of the VRA affected health outcomes by increasing perceived status threat. We first show that the VRA led white adults to perceive a rise in non-white status. Using survey data, we find suggestive evidence that white respondents in VRA covered areas were more likely to perceive that Black individuals – who comprised the largest share of non-white persons during the study era -- saw a relative improvement in their social and economic position after policy implementation. We find no such increase among non-white respondents. These survey findings can be interpreted in the context of studies which use the salience of group position and Republican support  -- which, since the early 1960s, has increasingly reflected racially conservative views (Kuziemko and Washington 2018) – as markers of status threat (Craig and Richeson 2014; Craig et al. 2018; Mutz 2018; Siddiqi et al. 2019).

We then show that mortality increases among white adults were larger in counties with below median levels of high school completion. This pattern is consistent with studies demonstrating how threats to their social status or economic position are more salient for individuals with lower levels of education and resources (Kuziemko et al. 2014; Reeping et al. 2024). Similarly, we find that increases in mortality among older non-white men are driven by the same less educated counties. These findings are consistent with a long-history of retaliatory responses limiting non-white men's—particularly Black men's—social and economic opportunities (Chetty et al. 2020; Derenoncourt 2022; Schechtl and O'Brien 2024; Luck 2025) and a growing literature linking status threat to retaliation against marginalized groups (Craig and Richeson 2014; Wetts and Willer 2018; Parker and Lavine 2025).

Our findings have several implications. First, they illustrate how policies to expand political power to disenfranchised groups may have nuanced effects on well-being. In doing so, our results meaningfully add to the growing literature on the mixed health consequences of the VRA for newly enfranchised groups. Prior work simultaneously documents both positive effects, such as increases in voter turnout, improved political representation and accountability, increases in public goods provision, improvements in economic outcomes, and reductions in political violence (Cascio and Washington 2014; Davidson and Grofman 2021; Bernini, Facchini, and Testa 2023; Lacroix 2023; Bernini et al. 2024; Aneja and Avenancio-Leon 2025), alongside outcomes that ostensibly reduce well-being among minority groups, including shifts to conservative politicians and increases in incarceration rates (Ang 2019; Eubank and Fresh 2022; Bernini, Facchini, Tabellini, et al. 2023). They add important nuance to the larger literature on the effects of political power on health and human capital outcomes (e.g., Miller 2008; Beaman et al. 2012; Bhalotra et al. 2023; Lynch 2023), which generally has found that extensions of power to marginalized groups either improve health across the board or improve outcomes among marginalized groups without resulting in negative effects for other groups. The difference between the findings in these studies and ours may be explained by unique salience of the Voting Rights Act – which was shaped by the broader Civil Rights activism, rising economic inequality, and falling social mobility that characterized the era in which it was implemented and extended (Donohue and Heckman 1991; Chetty et al. 2017) – and it is resulting impacts on perceived threats to social status.

Second, our findings underscore the importance of group identity and status threat for population health. To date, the literature on status threat has been descriptive, with no clear consensus on how to measure status threat and causally link those measures to health outcomes (Efird et al. 2023). Our study advances this literature by identifying a plausibly exogenous shock to status perceptions, examining the impacts of that shock on both targeted and non-targeted populations, and leveraging dynamic treatment effects and multiple data sources to measure and delineate the potential importance of status threat for health.

Finally, our findings are relevant to contemporary mortality patterns in the United States. A large literature has documented stagnant -- and eventually falling -- life expectancies among non-Hispanic white individuals without college degrees since the 1990s. This has occurred alongside rising life expectancies for non-Hispanic Black individuals without college degrees during much of this period (Case and Deaton 2021; Schwandt et al. 2021). These patterns have manifest even though both groups have faced similar economic challenges and many social and clinical risk factors for mortality have remained higher for Black Americans. Our findings support the emerging hypothesis that relative shifts in group economic status and political power may explain these discordant patterns (Siddiqi et al. 2019).

## I. Setting and Data

*A. Setting – U.S. Voting Rights Act and 1975 Extension*

The 1965 Voting Rights Act was a cornerstone policy of the Civil Rights Era. The VRA outlawed discriminatory voting practices, such as the use of literacy tests and established pre-clearance, in which any changes to voting laws or procedures required assessment and approval by the U.S. Department of Justice (Davidson and Grofman 2021; Bernini et al. 2024). The VRA, which initially applied to Southeastern states (Louisiana, Arkansas, Mississippi, Alabama, Georgia, Virginia, South Carolina, and parts of North Carolina), was extended twice, in 1970 and 1975.

The 1975 extension, signed into law on August 5$^{th}$ of that year, focused on expanding the reach of federal pre-clearance to address more subtle forms of voter discrimination, particularly among linguistic minority groups such as Hispanic, Asian, or Native American populations (Fraga 2017; Ang 2019). The extension specifically focused on states or political sub-divisions with high shares of a single language minority group (>5% of the voting age population) and where voter turnout was <50% in the 1972 election (U.S. Dept of Justice 2015). Targeted jurisdictions were subject to an expanded definition of what constituted a "test or device" that restricted access to voting -- which for the 1975 extension additionally required that states or political subdivisions provide election materials (e.g., instructions or ballots) in multiple languages – in addition to the stipulations of the original 1965 Act. Altogether, these stipulations ultimately led to coverage of three entire states -- Texas, Arizona, and Alaska – as well as isolated counties in California, Florida, Michigan, New York, North Carolina, and South Dakota (U.S. Dept of Justice 2015).[2]

A growing literature has documented increases in non-white voter turnout and political representation because of the original VRA as well as the 1975 extension (Bernini et al. 2025). In addition, studies have documented redistribution of state resources to predominantly non-white counties (Cascio and Washington 2014), improvements in labor market outcomes and declines in the racial wage gap (Aneja and Avenancio-Leon 2025), and reductions in political violence against marginalized groups (Lacroix 2023). This literature has also documented evidence of backlash against non-white groups in covered jurisdictions, some of which started immediately. Evidence of backlash includes increased mobilization and support for Republican politicians by white voters, reductions in overall public goods in predominantly non-white areas, and, over a longer time frame, increased rates of incarceration of Black individuals (Ang 2019; Eubank and Fresh 2022; Chaudhry 2025; Bernini et al. 2025). The pushback persisted for several decades, ultimately culminating with the U.S. Supreme Court striking down preclearance in their 2013 *Shelby v. Holder* decision. Thereafter, covered jurisdictions immediately reconstituted a range of barriers to curtail access to voting for minority groups and recent work has found evidence of erosion of Black economic outcomes soon thereafter (Aneja and Avenancio-León 2019).

---

[2] In this case of Texas, the Chicano political movement in the mid 1960s and early 1970s revealed the presence of barriers to voting among minorities that, while not explicitly a "test or device," rivaled the Deep South in their stringency. This recognition helped set the stage for the 1975 VRA extension and resulted in the entirety of Texas falling under coverage under the Act (Hunter 1976).

We study the 1975 extension of the VRA for several reasons. First, coverage came as a surprise to many jurisdictions, given both the specific changes in the formula used to determine coverage and the fact that the data to calculate the formula were not available until just prior to implementation (U.S. Dept of Justice 2015; Ang 2019). This surprise element provides a strong basis for identification. Second, the 1975 extension came a decade after the original VRA, allowing for accumulation and widespread diffusion of knowledge about its effects on political power and thus, along with the surprise element of coverage, increasing its salience to newly covered jurisdictions.[3] Ang (2019) documents evidence of this salience, finding evidence of immediate backlash against the extension in the form of local and state challenges to the law as well as large increases in newspaper coverage of the VRA. This enhanced salience likely increases our power to detect the impacts of the VRA on mortality, an outcome that is "downstream" to what has thus far been studied in the economics and political science literature.

Finally, unlike the original 1965 implementation of the VRA, the 1975 extension occurred after the height of the Civil Rights and War on Poverty Era, which saw a flurry of policy changes – e.g., the introduction of Medicare and Medicaid, hospital desegregation, widespread rollout of Community Health Centers and Food Stamps, and increases in minimum

---

[3] For example, the authors of a 1975 report by the U.S. Civil Rights Commission – available well in advance of the passage of the 1975 VRA extension– noted that "The Voting Rights Act has contributed substantially to the marked increase of all forms of minority political participation in the last 10 years," going on to provide detailed examples of success as well as elements of potential backlash. Thus, it is likely that the positive impacts of the VRA on minority status in the U.S. were well known at the time of the 1975 expansion (U.S. Commission on Civil Rights 1975).

wages – that have been linked to improvements in health.[4] These correlated policies – and the broader environment in which they occurred -- make it difficult to isolate the causal impacts of the 1965 VRA on health. Analyzing the 1975 extension allows us to circumvent these challenges.[5]

*B. Main data and sample*

Our primary outcomes are race-by-age group specific mortality rates, which we calculate at the county-year level using the universe of U.S. Vital Statistics death certificates for the period 1970 to 1988 and population denominators from the U.S. Census Bureau.[6] We start at 1970 given that estimates of bridged race-age-sex population denominators became available that year and end at 1988 given changes in how race and ethnicity were categorized in death certificates starting in 1989.[7] Because death certificates from this era only capture white, Black, and other as racial categories, we compute mortality rates for "white" versus "non-white." By this classification, decedents of Hispanic ethnicity (the group that comprised the bulk of language minorities in the United States during the study era) may have been assigned either white or non-white race in death certificates.[8] We address potential implications for our estimates in robustness checks below. Following Goodman-Bacon and Bailey (2015), we focus on children under age 5, 5–19, 20–49, and 50+ year-olds. Within each race-age group, we winsorize mortality rates at the 95th percentile to account for implausible values.

---

[4] See for example, Almond et al. (2007) on hospital desegregation; Bailey and Goodman-Bacon (2015) on Community Health Centers; Hoynes et al (2016) on Food Stamps; Goodman-Bacon (2018) on Medicaid; and Musen 2023) on minimum wages.
[5] Rushovich et al (2024) examine the impacts of the 1965 VRA on infant mortality, adjusting for some of these potential confounders. They find statistically significant relative reductions in death rates for Black infants and non-significant relative increases for White infants. They did not examine other age groups.
[6] Specifically, we use publicly available NCHS multiple cause of death files from 1970-1988 and county intercensal population estimates from the U.S. Census Bureau to calculated race stratified (white, non-white) county-level all-cause mortality rates per 100,000 population. We assign decedents to county using the county of residence recorded in the death certificates. We age standardize mortality rates to the 1970 U.S. population. Following Goodman-Bacon and Bailey (2015), we calculated cause-specific age-adjusted mortality rates for each race group per 100,000 population for cardiovascular disease, cerebrovascular disease, cancer, pulmonary disease, infectious disease, and external causes (see Table A1). We also calculated overall and cause-specific rates by gender.
[7] The main change was the introduction of a field to record ethnicity (i.e., Hispanic versus non-Hispanic): see https://www.cdc.gov/nchs/hus/sources-definitions/life-expectancy.htm for details. The change may have shifted assignment of decedents to racial categories, making it difficult to compare outcomes for white versus non-white groups before and after 1988.
[8] The 1970 Census recorded Hispanic ethnicity. The share of the population reporting Hispanic ethnicity was 4.5%. At the county level, the correlation between the share of adults of Hispanic ethnicity and the share of individuals belonging to the single largest linguistic minority group in 1970 was 0.82. In contrast, the correlation between Hispanic shares and the non-white population share 0.07.

To capture counties covered by the 1975 VRA extension we use federal data sources as well as the replication archive from Ang (2019). We exclude from our sample all counties that were covered by the VRA prior to 1975, as they cannot serve as credible controls for the 1975-covered counties (Goodman-Bacon 2021). We also excluded counties in Alaska and Hawaii, given challenges in calculating accurate mortality rates. After restricting the dataset to counties that, for specific age-race groups, had no missing data on mortality, non-zero population sizes, and availability of key covariates, our final sample included 256 covered and 1,862 uncovered counties (**Figure 1**)[9]. Covered counties were generally smaller and had higher shares of non-white and linguistic minority populations (**Table 1**).

## II. Research Design

Our core research design is a difference-in-differences approach that compares differences in mortality for each race-age group before and after VRA implementation in counties covered by the 1975 VRA versus those that were not covered. Our general approach is captured in the following equation:

(1) $Y_{it} = \alpha \times Post_t \times CoveredVRA_i + \mu_i + \theta_t + e_{it}$

where $Y_{it}$ represents the mortality rate for the race-age group of interest in county *i* and year *t* and $\mu_i$ and $\theta_t$ are county and time fixed effects. $CoveredVRA_i$ is a binary term = 1 if the county was covered by the Voting Rights Act, and $Post_t$ is equal to 1 after VRA implementation (the main effects of these variables are subsumed by the county and time fixed effects, respectively). We consider 1976 and onward as the post-intervention, as the policy was passed in late 1975. The coefficient $\alpha$ captures the difference-in-difference estimate on mortality, reflecting the relative change in mortality before and after 1976 in covered versus uncovered counties. Because mortality was generally declining through the study period (Figure A1), a positive estimate on $\alpha$ would reflect a slowdown in this secular decline in covered versus uncovered counties and a negative $\alpha$ would reflect a larger decrease.

---

[9] These are the maximum number of included counties and reflect sample sizes for white adults ages 50+ and above. For some age groups, like white children ages 5-19, sample sizes will be smaller given a higher prevalence of zeros in calculated mortality rates. Similarly, the sample of counties we use for analysis of non-white mortality is smaller as well, given the smaller sample size of this population in many areas.

In addition to obtaining average treatment effects, we also assess dynamic impacts, using the following event study specification:

(2) $Y_{it} = \sum_{k=-6}^{k=13} \boldsymbol{\delta_k} \times \mathbf{1}[VRA_{it} = k] + \mu_i + \theta_t + e_{it}$

Here, the term $\mathbf{1}[VRA_{it} = k]$ is an indicator that equals 1 if unit $i$ is $k$ years relative to 1976, the first full year the VRA extension was in place (for uncovered counties it is always equal to zero). The coefficients $\boldsymbol{\delta_t}$ capture the evolution of treatment effects over time. The advantage of the event study in our context (beyond able to assess violations of the parallel trends assumption) is that dynamic effects can help reveal underlying mechanisms. Specifically, impacts that accrue immediately may better reflect affective or psychological mechanisms than material mechanisms, since material conditions (for example, shifts in public spending or changes in household economic circumstances) generally take time to materialize (Boen et al. 2025). For example, changes in the provision of public goods (e.g., schools or health spending) due to the 1975 VRA extension would likely not appear until the 1978 fiscal year at the earliest given budget and election cycles with detectable mortality impacts surfacing even later.

Despite the surprise element of the VRA extension and visual inspection of raw trends in mortality suggesting otherwise (Figure A1), baseline differences between covered and non-covered counties (**Table 1**) indicate the possibility of unobserved violations of the parallel-trends assumption. Prior work on the VRA has addressed this possibility in several ways, including using regression discontinuity or difference-in-discontinuities designs; difference-in-differences designs in which parallel trends are assumed to hold conditional on a range of covariates; or triple differences designs.

However, these approaches create challenges for our specific research question. For example, the geographic regression discontinuity approach implemented by Bernini et al (2025), which focuses on bordering VRA and non-VRA counties, would greatly reduce the statistical power for rare outcomes that may fluctuate from year to year, like mortality. A similar problem applies to using discontinuities generated by elements of the 1975 VRA coverage formula (as did Ang 2019 in a robustness check). Introducing covariates to address potential violation of the parallel trends assumption is a widely used option but may limit available variation – again, with implications for power[10] – and allow for considerable researcher degrees of freedom (Roth 2022). Triple difference models, used by Cascio and Washington (2014) and Bernini et al (2023), allow the researcher to make a single, potentially weaker, parallel trends assumption (Olden and Moen 2022). However, this approach requires the researchers to either specify an untreated group (typically white individuals) or focus on the evolution in disparities in an outcome across two groups. In contrast, our goal is to assess treatment effects for specific population group in covered areas, including those not directly targeted by the intervention but potentially still affected by it.

To circumvent these challenges, we adopt a synthetic difference-in-differences (SDID) approach, which selects a weighted set of control units (based on unit and time weights) that minimize pre-intervention trends in the outcome (Arkhangelsky et al. 2021). This design allows us to make a weaker parallel trends assumption while using data for all sample counties (in contrast to border regression discontinuity designs) and without needing to assume an untreated population group (in contrast to triple differences model). SDID also limits researcher degrees of freedom by balancing linear unobserved factors rather than using covariates or sample restrictions to achieve conditional parallel trends and has the benefit of increased efficiency relative to TWFE in many settings (Dench et al. 2024). We compute standard errors accounting for clustering at the county-level using the block bootstrap approach recommended by Arkhangelsky et al. (2021).

It is important to note that the weights used in SDID are not population weights. Consequently, with SDID we are recovering the average treatment effect of the VRA on mortality rates in affected counties. The use of population weights would recover a different parameter, namely the effects of the VRA on mortality among people living in covered jurisdictions (Baker et al., Forthcoming). While it is common to use population weights in studies of area-level mortality, doing so presents additional challenges in our context. The main challenge is that some covered jurisdictions (e.g., Bronx County in New York City, Harris

---

[10] For example, Ang (2019) includes state-year fixed effects in his main specification to study the effects of the 1975 VRA extension on enfranchisement. The resulting estimates are identified off *within state* exposure to the VRA – the state-year fixed effects absorb any impacts from fully covered states like Texas and Arizona, whose counties comprise the vast bulk of covered jurisdictions. Such restrictions may not restrict power for outcomes like voter turnout but would certainly do so for rarer outcomes like mortality.

County (Houston), and Dallas County) have substantially larger populations than the median covered jurisdiction in the sample. As a result, population weighted treatment effects will be skewed markedly toward large cities. Such estimates would potentially be misleading, given that large cities differed substantially in socioeconomic and demographic terms and faced unique mortality challenges during the study period, e.g., significant burden of diseases from HIV, external causes like homicide, and deindustrialization, which leading to worsening general health in affected communities (Downs 1997; Geronimus et al. 1999; Wilson 2011). We further examine the issue of population weighting in Section III.B.

In all models, we account for time-varying differential exposure to the oil and natural gas boom and bust, which began in the early 1970s and crested in the early 1980s. This shock, particularly concentrated in Texas, Louisiana, and Oklahoma, may have had independent impacts on county-level mortality through changes in household resources, health care access and spending, population composition, and pollution (Acemoglu et al. 2013; Boslett and Hill 2022; Jacobsen et al. 2023; Moorthy and Shaloka 2025). To address this possibility, we include as a covariate a measure of exposure to the resource shock. We follow Allcott and Keniston (2018), using baseline natural resource endowments with annual global oil prices to create county-year estimates of exposure to the oil and gas boom.

In robustness checks, we also estimate specifications where we (1) do not include any covariates; (2) additionally adjust for the contemporaneous adoption of potentially confounding policies and programs in Arizona and Texas, such as the presence of a Community Health Center (see Bailey and Goodman-Bacon 2015) and implementation of the Food Stamp program (see Hoynes et al 2016); and (3) restrict the sample to only those counties where, for each age-race group, there is a recorded mortality event during the study period.[11] Given the possibility that both the non-white and white groups include individuals of Hispanic ethnicity, who may have also benefitted from the VRA given the explicit goal of extending the franchise to language minorities, we also estimate specifications where we restrict the sample to counties with low shares of language minorities in 1970.[12]

---

[11] This is relevant for younger non-white groups. For example, for non-white children ages 5-19 this exclusion reduces the number of included counties from 2,321 to 1,144. Most of this reduction is driven by counties not covered by the VRA. We do not expect these exclusions to change the results substantially, since SDID incorporates unit fixed effects (and the lack of effective variation in the outcome would effectively exclude counties without any mortality events given these fixed effects).

**III. Mortality Impacts**

*A. Core results*

SDID estimates of the impacts of the VRA extension on mortality by race-age group from our core specification are provided in **Table 2** (columns 1 and 5).[13] We find evidence of larger reductions in mortality for white and non-white children under the age of 5 years in counties covered by the VRA compared to mortality trends in the synthetic control group. The impacts – precisely estimated for both groups -- are larger for non-white children, who experienced an 80.8 deaths per 100,000 relative decrease in mortality due to the VRA (13.2% of the pre-VRA mean in this group) compared to a 21.1 deaths per 100,000 decline for white children (6.4% of the pre-VRA mean).

For children ages 5-19 and adults ages 20-49 we find opposing impacts of the VRA for non-white and white persons. For the non-white persons, we find precisely estimated relative declines in mortality of 2.2 deaths per 100,000 (a 5.8% decline relative to the pre-VRA mean) for 5–19 year olds and 30.5 deaths per 100,000 (an 8.3% decline relative to the pre-VRA mean) for 20-49 year olds. For white persons, we find evidence of increases in mortality of 4.9% and 2.4% for 5–19 year olds and 20-49 year olds relative to pre-intervention means for these age groups, although these are imprecisely estimated in our preferred specification.

For non-white adults over age 50, we find null estimates with standard errors over three times as large as the point estimates. These large standard errors are consistent with evidence of potential heterogeneous treatment effects of the VRA within this group, which we document in Section III.C and Section IV.D. For white adults over 50, we find a precisely estimated relative increase in mortality of 98.5 deaths per 100,000 in VRA vs non-VRA covered areas – a 3.4% increase compared to pre-intervention mean.

Event study estimates, presented in **Figure 2**, illustrate differences in the dynamics of treatment effects by age and race group. For all groups, we find no evidence of violation of parallel trends, suggesting that the SDID approach to reweight and match on pre-intervention trends was successful. For children under 5, we find evidence of gradual relative declines in mortality after implementation of the VRA extension. This same pattern can be seen for non-white older children and 20-49 year old non-white adults. In contrast, for older white children and adults, we find evidence of relative increases in mortality that occur in the first year of VRA implementation. For white adults, these year 1 estimates are precisely estimated. For white adults above the age of 50, they are sustained in magnitude and precision throughout the post-intervention period.

---

[12] Specifically, we restrict the sample to counties where the share of language minorities relative to the total population was 1.7% or less (corresponding to the cutoff for the bottom quartile of counties covered by the VRA; other splits produce similar results).
[13] Estimates of SDID unit weights are mapped in Figure A3 and estimates of time weights are provided in Table A2.

*B. Robustness Checks*

Our substantive findings (i.e., signs and overall magnitudes) are unchanged in specifications without any covariates (**Table 2**, columns 2 and 7 for non-whites and whites, respectively) and in specifications where we additionally adjust for county-year timing of the introduction of community health centers and the food stamp program (columns 3 and 8).[14] They are also unchanged when we restrict to counties with non-zero mortality events (for that age-race group) during the study period (columns 4 and 9) and when we use TWFE to estimate Equation (1) instead of SDID (Table A3, Columns 1 vs. 2 and 6 vs. 7; Figure A2 presents TWFE event studies).

The substantive findings are similar for non-white individuals when restricting the sample to counties with low shares of language minorities (who during the study era typically identified as Hispanic in the 1970 census; Column 5). In some cases, the coefficients are even larger in magnitude than the base estimates for this group (Column 1). However, the results substantively change for the white sample (Column 10). For the under-5 population, restricting to low language minority share areas attenuates the estimated VRA-led relative decline in mortality by over half, and the estimate is no longer statistically significant. The estimates in this restricted sample are larger in magnitude for two out of the three remaining white age groups compared to the base estimate (Column 6).

---

[14] Point estimates for white children 5-19 and white adults 20-49 are large and statistically significant in models without covariates but decline in magnitude and significance once the oil shock covariate is included. This difference is attributable to a positive relationship between exposure to the oil shock and external causes of deaths (such as accidents and injuries) in these age groups.

Collectively, these results suggest that individuals with Hispanic ethnicity – who we cannot observe directly in the data – likely benefitted directly from the VRA extension. To the extent that individuals of Hispanic ethnicity were classified as white in death certificates, this would bias estimates of the impacts of the VRA extension among white individuals downward -- perhaps explaining much of the reduction we find for white children under age 5.

We next examine the robustness of the results to weighting by population using TWFE models. Following Solon et al. (2015), we compare weighted and unweighted estimates to assess potential treatment effect heterogeneity. We find evidence suggesting this is the case: TWFE estimates using age-race group population weights yield estimates that are smaller in magnitude than our SDID or unweighted TWFE estimates (Table A3, Column 3 and 7). For 5-19 year olds, the weighted TWFE estimates for non-white persons are close to zero and those for white persons flip sign. We show that differences in weighted and unweighted estimates are reconciled once heterogeneous treatment effects by county size are considered. Specifically, once we include interactions between a binary indicator of being in the top decile of county population in 1970 (to denote large urban areas) and all right-hand side variables in Equation (1), we find estimates of the average treatment effect in weighted TWFE move closer to our SDID and unweighted TWFE estimates (Table A3, Columns 4 and 8). Evidence of heterogeneous treatment effects supports our justification of not using population weights, given the unique threats to health faced by large, metropolitan areas during the study period (see Section II).[15]

Finally, we also estimate TWFE triple difference models – typically used in economics work on the VRA and on racial health disparities in general (e.g., Almond et al. 2007; Cascio and Washington 2014; Alsan and Wanamaker 2018; Bernini et al. 2025) – which focus on relative changes in mortality for each age group between non-white and white individuals. To do so, we pool data for both racial groups and interact all right-hand side terms in Equation (1) with a binary indicator for non-white race. This model includes county-race, year-race, and county-year fixed effects. The last set of fixed effects account for county-specific annual shocks that may affect mortality in both racial groups. As discussed before, the benefits of requiring a weaker parallel trends assumption and assessing the impacts of the intervention on racial disparities in mortality comes with the cost of not being assess specific impacts for the racial group used as the reference (which are absorbed by the county-year fixed effects). These concerns aside, we find robust, precisely estimated evidence of larger mortality declines for non-white compared to white persons across all age groups (Table A4).[16]

---

[15] It is also consistent with the possibility that treatment effects may vary with other attributes that are correlated population size, such as area level of education. We examine this possibility in Section IV.D when discussing mechanisms.

*C. Mortality patterns by gender and cause of death*

We next present SDID estimates disaggregated by gender (**Table 3**) and cause of death (**Table 4**; fully disaggregated results by both gender and cause are provided in Table A5). For cause-specific analyses, we follow Goodman-Bacon and Bailey (2015) and focus on leading causes of death among adults during the study era, including cardiovascular disease, stroke, cancer, accidental deaths, and infectious diseases (see Table A1 for further details).

Analyses by gender reveal substantial heterogeneity. Both in absolute magnitudes and relative to pre-intervention means, the estimated VRA-led declines in under-5 mortality are larger for males than females (**Table 3**, columns 1 versus 2 for non-white children and 3 versus 4 for white children). For white children ages 5-19, we find a precisely estimated increase in mortality among males only, representing a 9.6% increase above the pre-intervention mean. Most striking – and consistent with the large standard errors in the overall estimates in **Table 2** – are the opposing estimates for non-white men and women above the age of 50. For non-white women, we find a reduction in mortality of 121.5 deaths per 100,000 (4.8% of the VRA mean) in VRA versus non-VRA covered areas. For non-white men, we find suggestive evidence of an increase in mortality in VRA covered areas of 113.2 deaths per 100,000 (statistically significant at $p<0.10$). The 3.1% relative increase in 50+ non-white male mortality is similar to the relative increase experienced by white men in the same age group. In additional analyses, we show that these results are unlikely to be explained by exposure to information about the Tuskegee Study of Untreated Syphilis, whose public revelation in 1972 has been linked to increased medical mistrust, reduced health care utilization, and stagnation in secular declines in mortality among older Black men (Alsan and Wanamaker 2018).[17]

---

[16] Using the baseline mortality rates provided in **Table 2** as a benchmark, the estimates imply a 20% reduction in the racial mortality gap among children under-5; a 21% reduction for children ages 5-19; a 17% reduction among adults ages 20-49; and a 72% reduction among adults ages 50 and above.

[17] Specifically, we compare estimates for counties above and below the median of the share of migrants in 1940 coming from Alabama, one measure of exposure to Tuskegee used by Alsan and Wannaker (2018). Following their study, we construct this measure using the 1940 full count census. While the SDID estimates from these split

Turning to causes of death among adults (**Table 4**),[18] we find that the relative increases in mortality for white adults are driven primarily by increases in cardiovascular disease deaths (precisely estimated for both 20-49 year old and 50+ adults) and strokes (precisely estimated for 50+ adults). Cancer mortality declined for both groups.[19] For non-white adults ages 20-49, we find large decreases in mortality from accidental causes. However, neither in **Table 4** nor in the analysis by cause and gender (Table A5), do we find dominant causes of death that account for the mortality patterns in this group. This could be explained by a larger share of other causes of death driving the findings or well-known measurement errors in race, age, and cause of death attribution for non-white deaths during the study period.[20]

**IV. Mechanisms**

In this section, we investigate several potential explanations for our core findings. Any set of candidate mechanisms collectively needs to explain the differing average and dynamic effects of the VRA by race-age (or race-age-gender) group.

*A. Selection*

Differential changes in population composition across counties (through migration or fertility choices) on the basis of VRA coverage could explain some of our results depending on whether these shifts varied by attributes correlated with mortality risk and did so differentially by race and age-group. We begin by noting that changes in migration and fertility patterns are unlikely to occur quickly enough to explain the immediate relative shifts in mortality for older white children and adults seen in event studies (**Figure 2**).

With this in mind, we investigate selection more formally using data from the Decennial Censuses (obtained from Social Explorer, www.socialexplorer.com). We first begin by

---

samples are imprecisely estimated, the point estimate for counties with below median shares of migrants from Alabama is large and positive (implying a relative increase of 166 death per 100,000) while the estimates for counties above the median – who would ostensibly be more exposed to the revelation of the Tuskegee experiment -- were smaller in magnitude and negative (implying a relative decrease of 33 deaths per 100,000).

[18] We do not examine causes of death for children given the distribution of causes varies substantially from adults. For children under 5, leading causes of death include congenital anomalies, infectious diseases, and accidents. For children ages 5-19, accidental deaths are the leading cause of death.

[19] A challenge in interpreting this result is the possibility of competing risks, i.e., that increases in cardiovascular mortality mechanically reduced cancer mortality rates since both affect similar (generally older) populations (Honoré and Lleras-Muney 2006).

[20] See, for example, Warshauer and Monk (1978), Rokaw et al. (1990) and Elo and Preston (1997).

estimating versions of Equation (1) using logged overall population, logged white population, and logged non-white population estimates from the 1960-1990 censuses as outcomes in SDID regressions. The results of this exercise are presented in **Table 5** (columns 1-3). We find relative increases in overall population size of 9% in VRA covered counties versus uncovered counties, with the non-white population growing by 36% after the VRA compared to 3% for the white population.

For this population growth to explain our mortality findings, we would need to simultaneously find evidence of selective population growth driven by non-white individuals who are likely in better health and white adults who are likely to be in worse health. To assess this possibility, we investigate census data on population shares of individuals above age 25 by race group with a high school degree or above. This measure is available for comparable racial groups in the 1970 and 1980 censuses. We focus on this measure given that high school completion status was economically significant in this era, was correlated with health outcomes, and was unlikely to have been influenced by the implementation of the VRA extension in this age group. Given we have only one pre-VRA time point, we estimate TWFE versions of Equation (1). The results, presented in **Table 5** (columns 4 and 5), run counter to what we find with mortality: the share of adults with at least a high school degree dropped for non-whites and increased for whites. We conclude from these exercises that population selection is unlikely to explain the heterogeneous mortality impacts of the VRA, particularly among adults. At best, selection may explain some of the decline in mortality experienced by white children under age 5.[21]

*B. Material resources*

Another potential explanation of the difference in the sign in mortality impacts for non-white and white older children and adults may be due to redistribution of material resources from the latter group to the former. Redistribution can take the form of how public services are allocated as well as shifts in the labor market which may lead to changing household economic circumstances. This explanation may be consistent with the smaller relative decline mortality experienced by white children under age 5, as well.

---

[21] We estimate the extent to which selection may explain reductions in white under-5 mortality using the following back-of-the-envelope calculation. First, we regress mortality rates for this group on the share of adults with high school degrees or higher in the year 1970. We then multiply the coefficient estimate from this model against the average increase in high school graduation shares among white adults as a result of the VRA (from Table 5). Doing so implies a reduction in under-5 mortality of nearly 7 deaths per 100,000 (78*0.089). That is equivalent to a third of the estimated mortality reduction for this group (**Table 2**, Col 6) and 71% of the reduction implied by estimates derived from the sample of counties with low shares of linguistic minorities in 1970 (**Table 2**, Col 10).

The literature is mixed on this possibility. Examining public spending after the 1965 VRA, Cascio and Washington (2014) find evidence of within-state redistribution to counties with higher proportions of Black residents. In contrast, Chaudhry (2025) leverages cross-state variation in VRA coverages and finds overall reductions in revenues and spending in covered non-white counties, which would run opposite to the mortality patterns we find. Focusing on individual economic outcomes, Aneja and Avenacio-Leon (2025) find evidence of substantial increases in wages among Black workers along with modest decreases in wages among White workers after the implementation of the original VRA in 1965.

We investigate redistribution of resources as a potential explanation for our heterogeneous mortality findings in several ways. We begin by examining shifts in public spending. To do so we use data from the U.S. Census of Governments, which is fielded every five years, typically in years ending in 2 and 7. We use data from 1967, 1972, 1977, 1982, and 1987. The use of both 1967 and 1972 data provides us the minimum two pre-intervention periods needed to the estimate Equation (1) using SDID. Using this approach, we find evidence of an increases in (logged) direct expenditures per capita in VRA covered versus uncovered counties (**Table 6**, top panel). These increases are larger for counties above the median in terms of the 1970 non-white population share. However, in counties with lower non-white population shares, the point estimates also suggest growth in direct expenditures per capita in covered versus uncovered counties, though these are imprecisely estimated.

We next examine changes in per capita income, which we do in two complementary ways. First, we estimate Equation (1) using annual data on logged county per capita income (from the Bureau of Economic Analysis) for the overall sample and for counties above and below the median of the 1970 non-white population share. The SDID estimates suggest increases in per capita income in both groups of counties (**Table 6**, second panel), though these are slightly larger in above median non-white share counties versus below median non-white share counties. Second, to better assess shifts in household income by racial group, we use individual-level data from Current Population Survey Annual Social and Economic Supplements (CPS-ASEC, Flood et al. 2024). Because the CPS-ASEC does not identify counties below 100,000 residents, we define exposure at the state-level, with Texas and Arizona considered treated states and all other continental U.S. states (excluding those covered by the 1965 VRA) as controls.[22] We use data from the 1964-1967 and 1977-1988 rounds as the CPS did not identify individual states (and instead identified state groups) in the intervening period. We collapse the data down to the state-year-racial group (white vs. non-white) level to apply the SDID estimator to the following equation:

$$(3) \quad Y_{jt} = \alpha \times Post_t \times CoveredVRA_j + \boldsymbol{\beta} \times \boldsymbol{X_{jt}} + \mu_j + \theta_t + e_{jt}$$

where $j$ the state of residence, $t$ the survey year, $\boldsymbol{X_{jt}}$ is a vector of covariates (state-year averages of respondent age, gender, and level of education), and $CoveredVRA_j$ a binary indicator denoting whether the state in question was fully covered by the 1975 VRA extension. We adjust standard errors for clustering at the state level. We note that estimates from equation (3) may underestimate the impacts of the VRA given the existence of covered counties in some of the control states; they may also be more prone to bias from omitted variables, for example given the inability to adjust for local shocks like the oil boom and bust. Estimates for both the log and level of household income are larger in magnitude for non-white households than for white households (**Table 6**, bottom two panels). Nevertheless, the coefficients remain positive and precisely estimated for white households. We also examine changes in household income for the above 50 population and those with below national median educational attainment. The former is motivated by the fact that the estimated VRA-led relative mortality increases among white individuals is most robust and precise for this age group. The latter is motivated by the potential for heterogeneous effects by socioeconomic status (which we explore in Section IV.D). The estimates for both groups are substantively similar to what we find for the full sample: estimates are larger in magnitude for non-white individuals but are still positive and statistically significant for white individuals (see Table A6).

      Altogether, these findings are inconsistent with the idea that redistribution of resources to groups experiencing relative increases in power came at the expense of those experiencing relative decreases. In particular, the economic position of white individuals in VRA-covered

---

[22] We remove respondents from Alaska and Hawaii to be consistent with our mortality sample.

counties appears to have improved in absolute terms and is thus unlikely to account for the relative increase in mortality experienced by older children and adults in this population after the VRA.[23] Improving resources also cannot explain the relative increase in mortality we find for non-white men above age 50. However, it may explain at least part of the mortality decline experienced by non-white children and adults 20-49, as well as white children under age 5.

*C. Health care crowd-out*

Even without reductions in material resources, mortality may have increased among white individuals in VRA-covered counties if increases in economic and political power for non-white individuals resulted in greater utilization of scarce health care resources. That is, increased utilization by a previously marginalized group may crowd-out utilization in other groups. Such an explanation is unlikely given historical and contemporary evidence that health care resource constraints typically disadvantage minority groups.[24]

Nevertheless, to more explicitly address this possibility, we examined whether the effects of the VRA on mortality patterns varied in counties with above and below median shares of hospitals per capita. The SDID estimates of Equation (1) after splitting the sample in this manner are presented in **Table 7**. Focusing on white individuals, the difference in the coefficients for older children and 20-49 year old adults may be consistent with crowd-out. However, the differences for children under age 5 and adults over age 50 run counter to this explanation. We interpret this as at best mixed evidence of crowd-out. While crowd-out may potentially account for some patterns in the data, it cannot explain the mortality patterns seen among older adults.

*D. Status threat*

Having presented evidence inconsistent with selection, differential shifts in material resources, and health care crowd-out as key explanations for the differential effects on mortality we find by race and age group, we now interrogate intergroup – or status -- threat resulting from VRA-driven increases in political power among non-whites relative to whites as a potential mechanism. A growing literature in the social sciences and psychology has documented how increases in minority group status creates a perception of status threat among majority groups (Blumer 1958; Blalock 1967; Craig and Richeson 2014; Mutz 2018; Craig et al. 2018; Metzl 2019; Stewart and Willer 2022; Efird et al. 2023; Parker and Lavine 2025). The bulk of this literature has examined how this sense of threat shapes political support and leads to punitive action or retaliation by majority groups against minority groups. A nascent literature has begun to link status threat to health outcomes, positing that perceptions of threat can harm health directly through psychosocial stress pathways and more indirectly through adoption of policy choices to retaliate against the minority group while inadvertently affecting the majority group (Siddiqi et al. 2019; Efird et al. 2023; Metzl 2024).

---

[23] That the circumstances for white individuals improved less in relative terms compared to non-white individuals is potentially important for our proposed explanation of status threat, and we will return to it below in Section III.D.
[24] For example, Almond, Chay, and Greenstone (2007) find no evidence of increases in white infant mortality after hospital desegregation in the 1960s, suggesting crowd-out effects were at best minimal. Similarly, early 20th century increases in hospitals per capita – which alleviated access barriers and ostensibly reduced the risk of crowd-out -- improved health outcomes more for Black infants than white infants (Hollingsworth et al. 2024). For a contemporary example, Singh and Venkataramani (2025) find that that mortality for Black, but not white, patients increases when hospitals reach capacity, a situation where crowd-out is more likely to occur.

Several of the findings presented thus far are consistent with perceptions of status threat being a key underlying mechanism. The first regards the dynamics of the relative increase in white mortality. The event studies in **Figure 2** demonstrate sharp, precisely estimated relative increases in mortality timed with the year of VRA implementation. These sudden increases are both opposite in sign and timing to the more gradual changes in mortality we document for non-white individuals. The dynamics of the mortality effects among white older children and adults is analogous to the sharp increases in Republican political support among white voters and widespread newspaper coverage of the 1975 VRA extension documented by Ang (2019), both of which are markers for backlash against the policy. In addition, these increases occur despite the absolute improvements in material resources among white households or in counties with higher white population shares. Instead, that the mortality increases for white groups occurs alongside a *relative* decline in material resources for white households compared to non-white households following the VRA (**Table 6**) is consistent with the hypothesis of status threat resulting from perceived disruptions to the economic status quo. Altogether, the sharp increases in mortality are consistent with immediately emerging psychosocial stressors -- such as status threat -- as a key driver of mortality among white older children and adults. That cardiovascular diseases and stroke – causes of death that have been linked to stress in the medical literature (e.g., Geronimus et al. 2006; Osborne et al. 2020) -- account for the bulk of these mortality increases further supports this contention.

Second, the relative increase in mortality experienced by older non-white males in VRA covered areas – a pattern that is distinct from all other non-white gender-age groups – is consistent with retaliation against a newly empowered minority group.[25] Such a gendered pattern of retaliation in ways that result in bodily harm is analogous to other historical episodes, for example lynching in the post-Reconstruction era (Beck and Tolnay 1990; Tolnay and Beck 1995; Williams et al. 2021) and increases in incarceration and murder rates limiting Black men's upward mobility after the second wave of the Great Migration (Derenoncourt 2022).

To further investigate the possibility of status threat as an underlying explanation for our mortality results, we conduct two additional sets of analyses. First, we directly examine perceptions of group position. To do so, we use data from the American National Election Studies (ANES), a series of in-person, representative surveys of the electorate conducted during congressional and presidential elections. This survey includes the question that queries group position for one, specific non-white group: "In the past few years we have heard a lot about improving the position of Black people in this country. How much real change do you think there has been in the position of Black people in this country: a lot, some, or not much at all?"[26] The responses to this question are coded as integers from 1-3, with 3 representing the stance that the position of Black people has changed "a lot." We also simultaneously analyze a binary measure of Republican party identification – which increasingly reflected racially conservative attitudes over the study period (Kuziemko and Washington 2018) – as an additional outcome. Because ANES data only include state identifiers, we estimate a version of the difference-in-differences model in Equation (3), which specifies exposure to the VRA at the state-level. Given the relatively small number of respondents per state and the fact that the ANES is not representative at the state level, we do not collapse the data down to the state-year-race group level to estimate the model using SDID as we did with the CPS. We instead use TWFE on individual level data (including respondent age and gender as covariates) and compute state-level clustered standard errors using both standard approaches and the wild cluster bootstrap-t approach (Cameron et al. 2008).[27]

---

[25] The smaller relative decline in mortality experienced by non-white men ages 20-49 compared to non-white women in the same age group is also consistent with this possibility.

[26] During our study period this question was fielded in 1970, 1972, 1976, 1984, 1986, and 1988. The question on party affiliation was fielded for all even years between 1970 and 1988.

[27] We use the wild cluster bootstrap-t approach for two reasons. First, after excluding states covered by the original implementation of the VRA, we are left with 35 clusters in the ANES data. Second, the cluster sizes are uneven.

The results are provided in **Table 8**. Among white respondents, perceptions of Black progress in America increased among those living in VRA-covered states versus uncovered status (these estimates are statistically significant when using conventional cluster corrected standard errors, though not so when using the wild cluster bootstrap-t approach). The corresponding estimate for non-white respondents is over two-thirds smaller in magnitude. In lockstep with these increased perceptions comes identification with the Republican party, which increased by 11 percentage points (42% of the pre-intervention mean) for white respondents in VRA-covered states, but decreased among non-white respondents. Together, these results suggest that VRA coverage made white adults more likely to perceive improvements in status for at least one non-white population: Black Americans. They also are consistent with findings by Bernini et al (2025), who find increased use of racial identifiers in local newspapers after the 1965 VRA and Ang (2019), who find greater and more sustained coverage of the 1975 VRA extension in counties with greater Republican support.

Our second set of analyses leverage the fact that threats to status should be more salient among individuals with lower socioeconomic status. For example, Kuziemko et al. (2014), though not studying racial threat per se, find that individuals lower in the income distribution are more averse to redistribution if it means that group below them achieve a higher position in the socioeconomic hierarchy than their own. Similarly, Reeping et al (2024) find that white adults with a high school degree or less are more likely to endorse political violence – a key manifestation of pervasive status threat – than white adults with higher education. Consequently, we would expect mortality increases to be more pronounced in counties with lower socioeconomic status. To test this hypothesis, we median split our sample by the share of white adults with a high school education in 1970 and estimate Equation (1) for each sample.[28]

The results, presented in in **Table 9**, accord with this hypothesis. For white adults, we find larger relative increases in mortality after the VRA in counties with below median baseline high school graduate shares – i.e., low socioeconomic status areas. These findings cannot be explained by changes in material resources among white individuals without a high school degree, which we find increased after the VRA (though relatively less so than it increased non-white adults; see Section IV.B and Table A6).

---

[28] We use area-level education given that birth certificates from the study era did not include level of education of the decedent.

For non-white older children and adults, we find that declines in mortality are also lower in counties with below median share of high school graduates. Strikingly, for non-white adults over the age of 50, we find evidence of decreasing in mortality after the VRA in counties above the median in white high school share completion and a small increase in counties with below median shares. This finding is driven by non-white men, for whom we find suggestive evidence of large VRA-led increases in mortality in counties where white high school education rates were low but not in more educated counties (**Table 10**). The findings for non-white men are consistent with health-limiting backlash against minority men in the face of shifts in relative status due to economic or political shocks (Beck and Tolnay 1990; Williams et al. 2021; Derenoncourt 2022). They are also consistent with emerging work demonstrating how conservative political shifts in the United States are more strongly and negatively correlated with health for Black men compared to women and white men (Luck 2025).

**V. Discussion and Conclusion**

We document substantial heterogeneity in the effects of expanding political power to minority groups on health. Using the 1975 extension of the Voting Rights Acts as a case study, we find generalized declines in mortality for most non-white individuals and, in some specifications, younger white children in VRA-covered counties compared to non-covered counties. However, we document evidence of relative increases in mortality among older white children and white adults. We also find suggestive evidence of increases in mortality for non-white men above the age of 50. These patterns – particularly the mortality increases – cannot be reconciled by non-random selection or shifts in material resources.

Instead, we find evidence consistent with status threat as a key mechanism by which enfranchisement of non-whites impacted mortality outcomes. In doing so, our findings provide further empirical support to a growing literature in social sciences and health on status threat (Siddiqi et al. 2019; Efird et al. 2023) and, more generally, the importance of psychological mechanisms in driving the health effects of social policies and political shocks (Torche and Rauf 2021; Boen et al. 2025). An advantage of our case study relative to the literature is that it represents a true, exogenous policy shock aiming to improve the status of a disenfranchised group, allowing us to address challenges with establishing causality and measuring status threat faced by other work on this subject. A limitation is that this is just one historical episode, and direct evidence of the mechanism of perceived threat – or how it was used to retaliate against non-white individuals beyond markers of political support -- is not available to us.

Nevertheless, our findings illustrate the potential importance of status threat in shaping key health trends in the United States. Consistent with the descriptive analysis by Siddiqi et al (2019), they offer an alternate explanation for the slow-down and increase in mortality among white adults over the last four decades as well as widening gradients in mortality by level of education. They also may help explain why life expectancy growth for white and Black men both stagnated during the 1950s-1970s relative to trends experienced by women (National Center for Health Statistics 2020). For White men, the implied mechanism would be status threat during an era of dramatic social change. For Black men, the mechanism would be retaliation in the face of social change by majority group, in addition to the effects of discrimination and mistreatment in health care (Alsan and Wanamaker 2018). Status threat may also help explain the growing link between Republican party affiliation and health outcomes in recent decades (Bor 2017; Wallace et al. 2023).

Finally, our findings of worsening mortality white adults and older children despite improvements in absolute economic circumstances highlights the importance of relative assessments of socioeconomic position and zero-sum thinking in shaping well-being (Chinoy et al. 2023; Lavetti et al. 2025). Our results invite the hypothesis that efforts to enfranchise marginalized groups may be more likely to result in backlash and result in negative outcomes in societies where inequality is perceived in this manner.

**Figure 1**. Sample counties

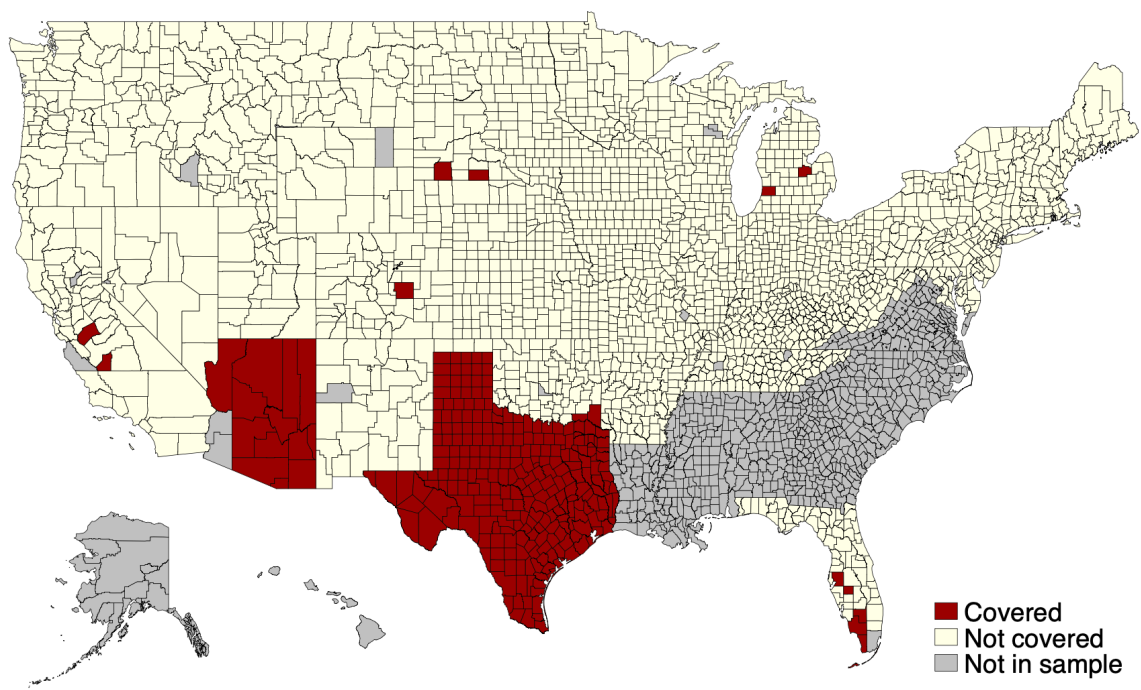

Notes: Figure plots counties covered by the 1975 VRA extension, counties not covered, and counties excluded from the sample. "Not in sample" includes all counties covered by the original 1965 VRA (since these areas were pre-treated), counties in Alaska and Hawaii (given difficulties in accurately calculating mortality rates), and any counties which recorded population (in a given age-race group) was equal to zero and for which there was missing data on outcomes and covariates. This map reflects the maximum number of counties included (and is based on mortality data for white individuals above age 50). For non-white groups, we end up excluding a larger number of counties because recorded population sizes are zero.

**Figure 2**. Dynamic impacts of the 1975 VRA Extension on mortality

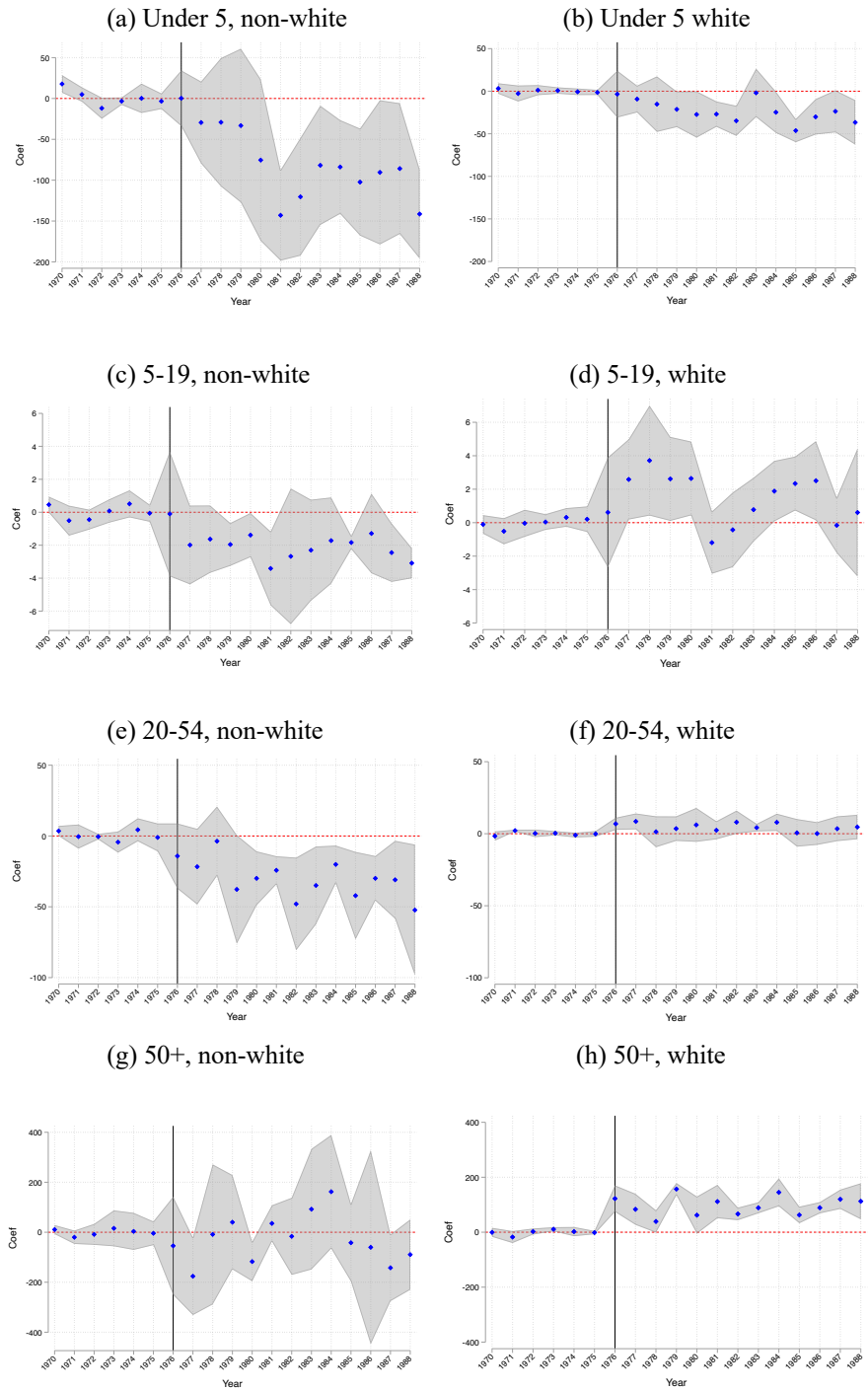

Notes: SDID event study estimates of the impacts of the 1975 VRA Extension on mortality by race-age group. 95% CIs (adjusted for clustering at the county-level) in parenthesis. All models adjust for time-varying exposure to the oil shock. Y-axes are scaled to be equivalent within each age group. See Table 2 notes for further details. Event studies from TWFE models are presented in Appendix Figure 3.

**Table 1.** Descriptive statistics

|  | Not Covered | | Covered | |
|---|---|---|---|---|
|  | Mean | SD | Mean | SD |
| Total Population | 22184.4 | 21702.2 | 18120.5 | 18761.5 |
| % White | 96.1 | 8.1 | 90.3 | 12.0 |
| % Non-White | 3.9 | 8.1 | 9.7 | 12.0 |
| % College | 6.9 | 3.3 | 7.5 | 3.4 |
| % Language Minority | 2.4 | 7.7 | 9.9 | 13.1 |
| Per Capita Income | 10205.5 | 2118.0 | 9851.7 | 2326.6 |
| % Low Income | 23.4 | 10.1 | 29.0 | 9.7 |
| % Employed | 50.9 | 6.2 | 50.6 | 5.5 |
| Hospitals per 100,00 capita | 8.0 | 8.7 | 11.2 | 11.9 |
| N | 1862 | | 256 | |

Notes: Baseline sample county characteristics by VRA coverage. % white, % non-white, % college (for individuals above age 25), per capita income, % low income, and % employed are all from the 1970 U.S. Census. Data on the population share of the largest language minority are for 1960 and obtained from Ang (2019). Data on hospitals per 100,000 persons are from the 1975 Area Resources Files.

**Table 2**. Overall impact of the 1975 VRA Extension on mortality

|  | Non-white | | | | | White | | | | |
|---|---|---|---|---|---|---|---|---|---|---|
|  | (1) | (2) | (3) | (4) | (5) | (6) | (7) | (8) | (9) | (10) |
| **Under 5** | -80.82*** | -78.18*** | -81.87*** | -84.32*** | -77.1*** | -21.11*** | -23.21*** | -21.42*** | -21.49*** | -9.82 |
|  | (17.91) | (15.78) | (18.28) | (17.75) | (26.80) | (5.80) | (5.47) | (8.05) | (7.42) | (14.0) |
|  |  |  | *608.3* |  |  |  |  | *326.7* |  |  |
| **5-19y** | -2.15*** | -1.99*** | -2.47*** | -2.60** | -3.70*** | 1.04 | 1.42** | 1.01 | 1.03 | 0.95 |
|  | (0.65) | (0.66) | (0.74) | (1.30) | (1.47) | (0.71) | (0.57) | (0.76) | (0.62)) | (1.51) |
|  |  |  | *36.7* |  |  |  |  | *21.4* |  |  |
| **20-49y** | -30.5*** | -30.02*** | -28.81*** | -24.49*** | -46.2*** | 3.69 | 4.45** | 3.68* | 3.60 | 4.87 |
|  | (8.05) | (8.43) | (8.82) | (8.84) | (15.5) | (2.56) | (1.88) | (2.16) | (2.59) | (3.74) |
|  |  |  | *368.3* |  |  |  |  | *151.4* |  |  |
| **50 and above** | -18.42 | -28.92 | -36.49 | -9.20 | -52.92 | 98.47*** | 96.82*** | 88.55*** | 98.47*** | 116.85*** |
|  | (58.99) | (46.12) | (54.91) | (63.74) | (93.36) | (16.17) | (15.02) | (15.38) | (17.55) | (23.93) |
|  |  |  | *3068.1* |  |  |  |  | *2904.7* |  |  |
| *Controls* | | | | | | | | | | |
| Oil and gas | X |  | X | X | X | X |  | X | X | X |
| CHCs |  |  | X |  |  |  |  | X |  |  |
| Food Stamps |  |  | X |  |  |  |  | X |  |  |
| Non-zero events |  |  |  | X |  |  |  |  | X | X |

Notes: Each column-panel presents synthetic difference in difference (SDID) estimates of the average treatment effects of VRA coverage on mortality for a given race-age group (Equation 1). Core estimates are presented in columns 1 and 6. Columns 2 and 7 present estimates without any covariates; Columns 3 and 8 add covariates for the presence of a Community Health Center and the timing of the introduction of Food Stamps; Columns 4 and 9 present estimates restricting the sample to only those counties where there was at least one recorded death in that particular age-race group during the study period; Columns 5 and 10 present estimates restricting the sample to counties with low (<1.7%) shares of linguistic minorities as recorded in the 1970 census. Block bootstrapped standard errors, accounting for clustering at the county-level, are in parenthesis. Baseline population mortality rates (per 100,000) for the main sample for each race-age group are in italics. Sample sizes (county-years) for the core specification are as follows: Under 5, non-white – 39,653; Under 5, white – 46,246; 5-19, non-white – 44,099; 5-19, white – 46,265; 20-49, non-white – 45,125; 20-49, white – 46,265; 50+ non-white – 43,643; 50+ white – 46,265. For the specifications in Columns 4 and 8, the sample sizes are as follows: Under 5, non-white – 26,697; Under 5, white – 46,108; 5-19, non-white – 21,736; 5-19, white – 45,334; 20-49, non-white – 33,060; 20-49, white – 46,246; 50+ non-white – 37,848; 50+ white – 46,265. Event study estimates for Equation 1 are presented in Figure 2. Estimates using two-way fixed effects are provided in Table A3.

*** - $p<0.01$, ** - $p<0.05$, * - $p<0.1$

**Table 3**. Impact of the 1975 VRA Extension on mortality by gender

|  | Non-white | | White | |
| --- | --- | --- | --- | --- |
|  | (1) | (2) | (3) | (4) |
|  | Men | Women | Men | Women |
| **Under 5** | -83.33*** | -52.20*** | -32.85*** | -7.01 |
|  | (15.06) | (14.35) | (9.54) | (7.663) |
|  | *665.3* | *534.4* | *368.2* | *281.8* |
| **5-19y** | -1.09* | -0.93** | 2.31*** | -0.21 |
|  | (0.64) | (0.40) | (0.83) | (0.78) |
|  | *41.2* | *24.9* | *24.1* | *18.3* |
| **20-49y** | -28.16** | -30.04*** | 4.481 | 2.737 |
|  | (11.07) | (7.332) | (4.366) | (2.605) |
|  | *482.1* | *265.2* | *187.2* | *116.6* |
| **50 and above** | 113.2* | -121.5** | 101.6*** | 88.97*** |
|  | (65.05) | (59.66) | (23.30) | (16.93) |
|  | *3682.0* | *2545.9* | *3521.0* | *2411.3* |

Notes: Each column-panel presents SDID estimates of the average treatment effects of VRA coverage on mortality for a given race-age-gender group. See Table 2 notes for further details. Baseline population mortality rates (per 100,000) for each race-age group are in italics. Block bootstrapped standard errors, accounting for clustering at the county-level, are in parenthesis.

*** - p<0.01, ** - p<0.05, * - p<0.1.

**Table 4**. Impact of the 1975 VRA Extension on mortality by cause of death

| | Non-white | | | | | White | | | | |
|---|---|---|---|---|---|---|---|---|---|---|
| | (1) | (2) | (3) | (4) | (5) | (6) | (7) | (8) | (9) | (10) |
| | Cardio. | Cerebrovasc. | Cancer | Infectious | Accidental | Cardio. | Cerebrovasc. | Cancer | Infectious | Accidental |
| **20-49y** | -2.06 | -0.93** | -3.08*** | -0.11* | -7.68** | 2.93*** | -0.04 | -2.06** | -0.02 | 2.09 |
| | (2.20) | (0.38) | (0.95) | (0.066) | (3.04) | (1.13) | (0.34) | (0.86) | (0.10) | (1.88) |
| | *78.6* | *20.2* | *42.3* | *5.9* | *70.7* | *45.0* | *7.9* | *31.9* | *1.84* | *48.8* |
| **50 and above** | 1.31 | -49.34*** | -11.08 | 1.77* | -2.33 | 77.09*** | 15.23*** | -19.8*** | 0.82 | -1.18 |
| | (30.73) | (12.75) | (12.38) | (1.02) | (2.49) | (12.11) | (4.79) | (4.91) | (0.65) | (1.93) |
| | *1305.4* | *399.4* | *523.8* | *34.8* | *90* | *1432.0* | *358* | *473.1* | *16.6* | *80.3* |

Notes: Each column-panel presents SDID estimates of the average treatment effects of VRA coverage on mortality from a specific cause for a given race-age. See Table 2 notes for further details regarding the regression specification and Table A1 for ICD-8 and -9 codes used to define causes of death. Baseline population mortality rates (per 100,000) for each race-age-cause of death group are in italics. We do not compute cause of death estimates for younger age groups given epidemiological differences in mortality risk factors differentiating adults and children (e.g., 5-19 year old deaths are predominantly driven by accidental causes while under 5 deaths were congenital and genetic anomalies, prematurity and low-birth weight related conditions, accidents and infections. Estimates by race-age group-gender are provided in Table A5. Block bootstrapped standard errors, accounting for clustering at the county-level, are in parenthesis.

*** - p<0.01, ** - p<0.05, * - p<0.1.

**Table 5**. VRA coverage, population growth, and composition

|  | (1) | (2) | (3) | (4) | (5) |
|---|---|---|---|---|---|
|  | ln(total pop) | ln(non-white pop) | ln(white pop) | Share HS and above, non-white | Share HS and above, white |
| **VRA Coverage** | 0.09*** | 0.36*** | 0.03** | -0.042*** | 0.089*** |
|  | (0.014) | (0.08) | (0.01) | (0.007) | (0.003) |

Notes: Columns (1)-(3) present SDID estimates of the impact of the 1975 VRA expansion on logged total, non-white, and white populations using data from the 1960-1990 U.S. Censuses. Columns (4) and (5) present TWFE estimates of the share of non-white and white adults above age 25 with a high school degree or above. Comparable data on these measures are available for all counties from two censuses. N refers to the total number of county-year observations. 1970-1980. For columns (1)-(3), we estimate standard errors using a block bootstrap procedure, to account for clustering at the county level. We compute cluster standard errors for the TWFE models in columns (4)-(5). Sample sizes (the number of county-year observations) are as follows: Col (1)-(3) = 9,072; Col (4) = 3,418; Col (5) = 4,536.

*** - $p<0.01$, ** - $p<0.05$, * - $p<0.1$.

**Table 6**. VRA coverage, area-level spending and per capita income, and household income

| | (1) | (2) |
|---|---|---|
| **County-level outcomes** | Above median non-white share | Below median non-white share |
| Ln(Total Direct Exp/Cap) | 0.11*** | 0.034 |
| | (0.038) | (0.028) |
| Ln(Area Income/Cap) | 0.0566*** | 0.0413*** |
| | (0.009) | (0.145) |
| **Household-level outcomes** | Non-white | White |
| Ln(Family Income) | 0.15 | 0.08*** |
| | (0.10) | (0.03) |
| Family Income | 4,030*** | 2,050** |
| | (1,080) | (1,040) |

Notes: Each column-row presents a separate regression. The first panel presents SDID estimates for total direct spending by all local governments within a given county and area personal income, both per capita. County-level spending data are from the U.S Census of Governments, which is fielded every 5 years (typically in years ending in 2 and 7). We use data from the 1967, 1972, 1977, 1982, and 1987 censuses. The use of pre-period data from both 1967 and 1972 provides us the minimum number of pre-intervention time points to implement SDID. County-level personal income estimates come from the Bureau of Economic Analysis and are available annually between 1970-1988. We estimate models separately for counties above and below the median of the 1970 non-white population share in VRA covered areas. Block bootstrapped standard errors, which adjust for clustering at the county level, are in parenthesis.

The second panel presents SDID estimates for logged and unlogged household income using data from the Current Population Survey Annual Social and Economic Supplement. Due to privacy concerns, individual state identifiers for smaller states (including one of the VRA covered states, Arizona) are not provided for the years 1968-1976; instead state-group identifiers are provided. Given this limitation, we use data from the 1964-1967 and 1977-1988 CPS-ASEC waves, which separately identify all states. Excluding all states that were previously covered by the VRA as well as Alaska and Hawaii, our final sample includes 43 states and the District of Columbia. We collapse data by state-year-racial group to facilitate estimation by SDID. Our final sample includes 748 state-year observations for each racial group. Block bootstrapped standard errors, which adjust for clustering at the state level, are in parenthesis. Estimates for specific subgroups – individuals above age 50 and lack of a high school degree – are presented in Table A6.

*** - $p<0.01$, ** - $p<0.05$, * - $p<0.1$.

**Table 7**. Heterogeneity in mortality impacts by baseline level of health care infrastructure

|  | Non-white | | White | |
| --- | --- | --- | --- | --- |
|  | (1) | (2) | (3) | (4) |
|  | Below median hosp.cap | Above median hosp/cap | Below median hosp.cap | Above median hosp/cap |
| **Under 5** | -104.0*** | -70.59*** | -26.39** | -15.65* |
|  | (25.79) | (18.89) | (10.89) | (8.92) |
|  | *599.9* | *573.3* | *319.9* | *358.1* |
| **5-19y** | -3.39*** | -1.99** | 1.56* | 0.76 |
|  | (1.25) | (0.89) | (0.81) | (0.84) |
|  | *36.8* | *23.2* | *21.3* | *20.6* |
| **20-49y** | -21.24 | -37.4*** | 8.41** | 1.49 |
|  | (13.6) | (9.80) | (3.41) | (2.93) |
|  | *366.0* | *350.3* | *148.7* | *153.7* |
| **50 and above** | 7.69 | -44.5 | 77.63** | 114.83*** |
|  | (69.1) | (73.6) | (30.65) | (20.16) |
|  | *3070.6* | *3320.7* | *2865.8* | *3068.1* |

Notes: Each column-panel presents SDID estimates of the average treatment effects of VRA coverage on mortality for a given race-age group, stratified by counties above and below the sample median of hospitals per capita prior to VRA implementation. Baseline population mortality rates (per 100,000) for each race-age group are in italics. We find no statistically significant differences in estimates for any race-age group in TWFE models that include the full set of interactions between a binary indicator of being above or below the median of hospitals per capita and all right-hand side variables

*** - p<0.01, ** - p<0.05, * - p<0.1.

**Table 8**. VRA coverage, impressions of minority progress, and political preferences

|  | Non-white | | White | |
|---|---|---|---|---|
|  | (1) | (2) | (3) | (4) |
|  | Change in position of Black people | Identify as Republican | Change in position of Black people | Identify as Republican |
| **VRA Coverage** | 0.034 | -0.012 | 0.13** | 0.098** |
|  | (0.053) | (0.028) | (0.053) | (0.039) |
|  | [0.64] | [0.67] | [0.20] | [0.21] |
|  | *2.32* | *0.06* | *2.48* | *0.26* |

Notes: TWFE estimates of the association between VRA coverage and perceptions of social and economic progress made by Black Americans and Republican party identification by race using data from the American National Election Studies (ANES). Data on perceived progress made by Black Americans comes from the question, "How much has the position of Blacks changed" with the answer choices being "not much at all" (assigned 1), "Some" (assigned 2), and "A lot" (assigned 3). Given that the ANES only identifies state of residence, VRA coverage is defined at the state-level (see Equation 2) and there are 35 states represented in the sample (excludes all states that were previously covered by the VRA as well as states where there were no survey respondents). All models include adjust for sex-age interactions and are estimated using sampling weights. Standard errors accounting for clustering at the state-level are provided in parenthesis. Wild cluster bootstrap t p-values are in square brackets. Baseline values of the outcomes are provided in italics. The sample size for Col (1) = 1,174; Col (2) = 2,343; Col (3) = 6,864; Col (4) = 13,376.

*** - p<0.01, ** - p<0.05, * - p<0.1.

**Table 9**. Heterogeneity in mortality impacts by baseline level of education

|  | Non-white | | White | |
|---|---|---|---|---|
|  | (1) | (2) | (3) | (4) |
|  | Below median HS | Above median HS | Below median HS | Above median HS |
| **Under 5** | -91.85*** | -94.72*** | -14.93** | -29.01** |
|  | (22.25) | (30.97) | (7.030) | (10.28) |
|  | *629.9* | *603.8* | *355.6* | *318.7* |
| **5-19y** | -1.387* | -4.488*** | 0.948 | -0.572 |
|  | (0.730) | (1.583) | (0.788) | (1.242) |
|  | *33.4* | *37.5* | *22.0* | *21.2* |
| **20-49y** | -30.04** | -50.03*** | 4.866* | 0.870 |
|  | (10.39) | (14.08) | (2.592) | (5.140) |
|  | *423.0* | *358.1* | *167.8* | *147.3* |
| **50 and above** | 3.23 | -169.0* | 105.8*** | 30.43 |
|  | (70.74) | (86.67) | (23.71) | (26.87) |
|  | *3467.1* | *2950.3* | *3085.7* | *2848.5* |

Notes: Each column-panel presents SDID estimates of the average treatment effects of VRA coverage on mortality for a given race-age group, stratified by counties above and below the sample median of the share of white adults with a high school education or greater. Baseline population mortality rates (per 100,000) for each race-age group are in italics. In TWFE models that include the full set of interactions between a binary indicator of being above or below the median of the share of HS graduates and all right-hand side variables, estimates for 5-19 y non-white and 50+ non-whites across education groups are statistically significant at p<0.10 and the differences in estimate for 50+ whites is statistically significant at p<0.05.

*** - p<0.01, ** - p<0.05, * - p<0.1.

**Table 10**. Heterogeneity in mortality impacts by gender and baseline level of education, non-white adults over age 50

|  | Men | | Women | |
|---|---|---|---|---|
|  | (1) | (2) | (3) | (4) |
|  | Below median HS | Above median HS | Below median HS | Above median HS |
| **VRA Coverage** | 146.9* | -66.79 | -124.2** | -147.8** |
|  | (77.0) | (97.9) | (57.9) | (72.1) |

Notes: Each column presents SDID estimates of the average treatment effects of VRA coverage on mortality among non-white adults over age 50 by gender, stratified by counties above and below the sample median of the share of white adults with a high school education or greater. Estimates from TWFE models that include the full set of interactions between a binary indicator of being above or below the median of the share of HS graduates and all right-hand side variables suggest that differences in VRA effects for non-white men above age 50 are statistically significant at p<0.05.

*** - p<0.01, ** - p<0.05, * - p<0.1.

Appendix for "Political Power and Mortality: Heterogeneous Effects of the U.S. Voting Rights Act"

**Figure A1**. Raw trends in mortality by race and age group

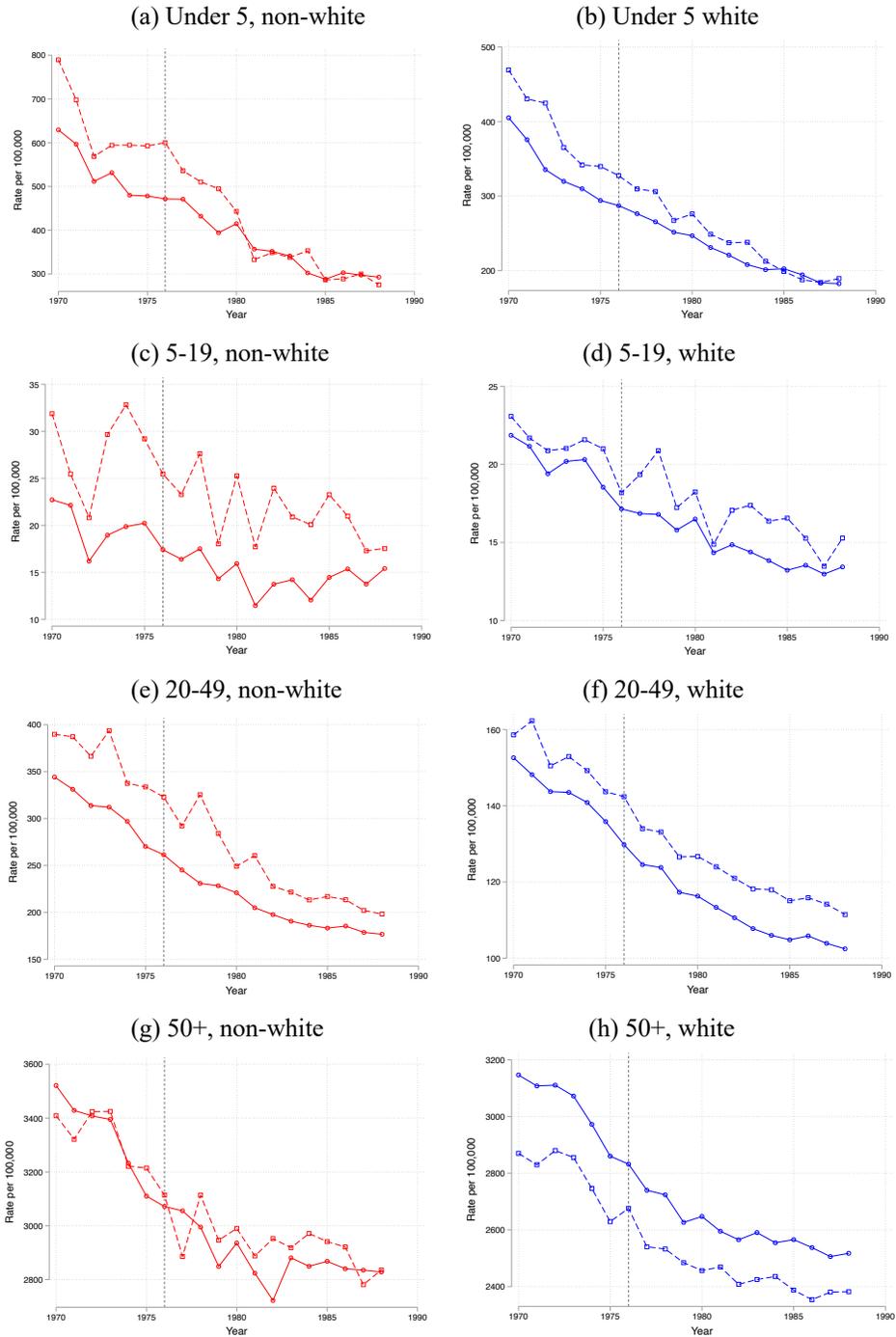

Notes: Raw trends in mortality by race and age group. Solid lines (with circle markers) mark trends for non-VRA counties and dashed lines (with square markers) mark trends for VRA covered counties. Mortality rates are weighted by population. The sample excludes counties in the top decile of 1970 population size given concerns around distinct mortality risks in larger cities during the study period (see main text for more details).

**Figure A2**. SDID Unit Weights

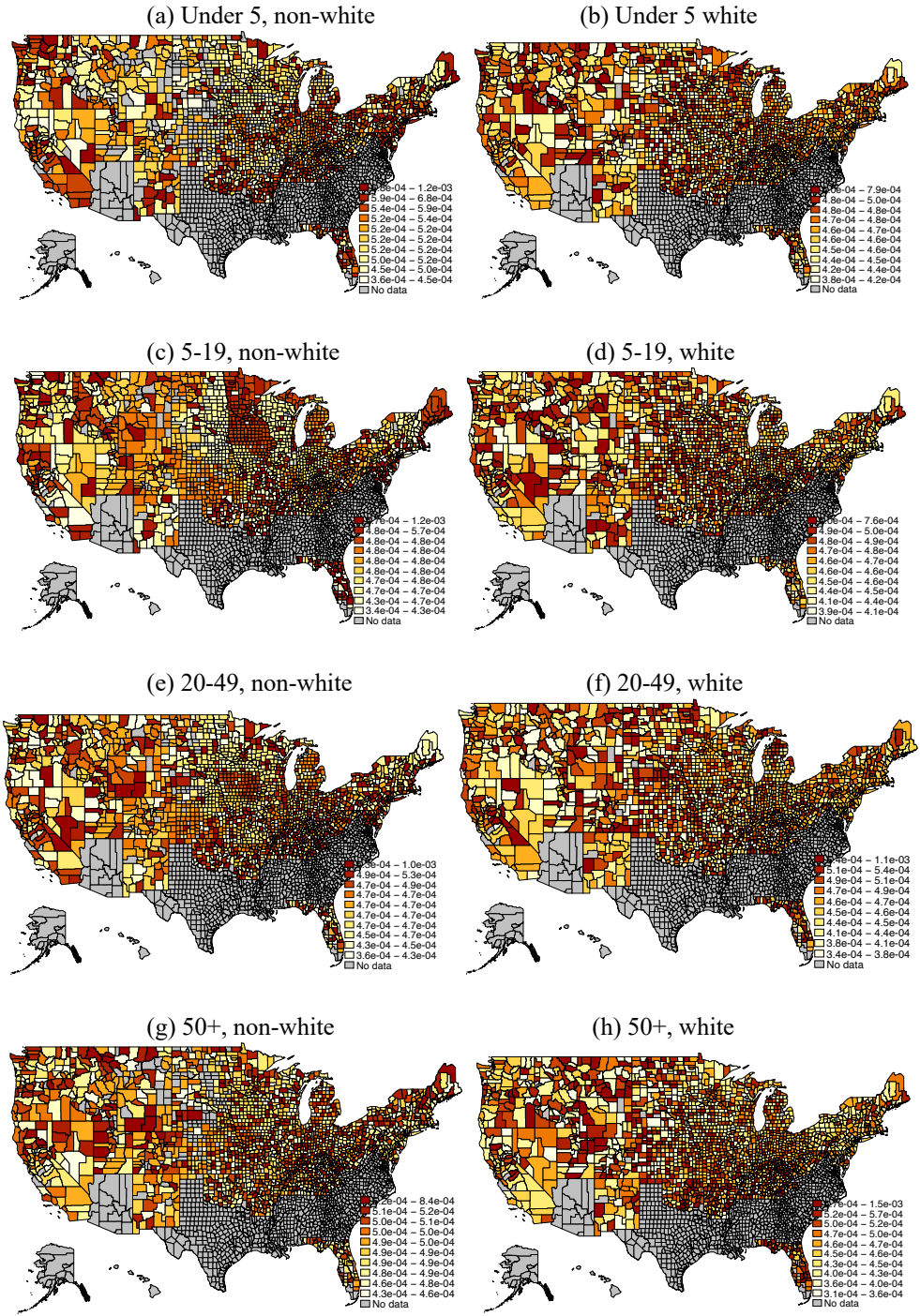

Notes: Estimates of SDID unit weights for the donor pool of control counties (non-grey areas) from Equation (1).

**Figure A3**. TWFE event studies

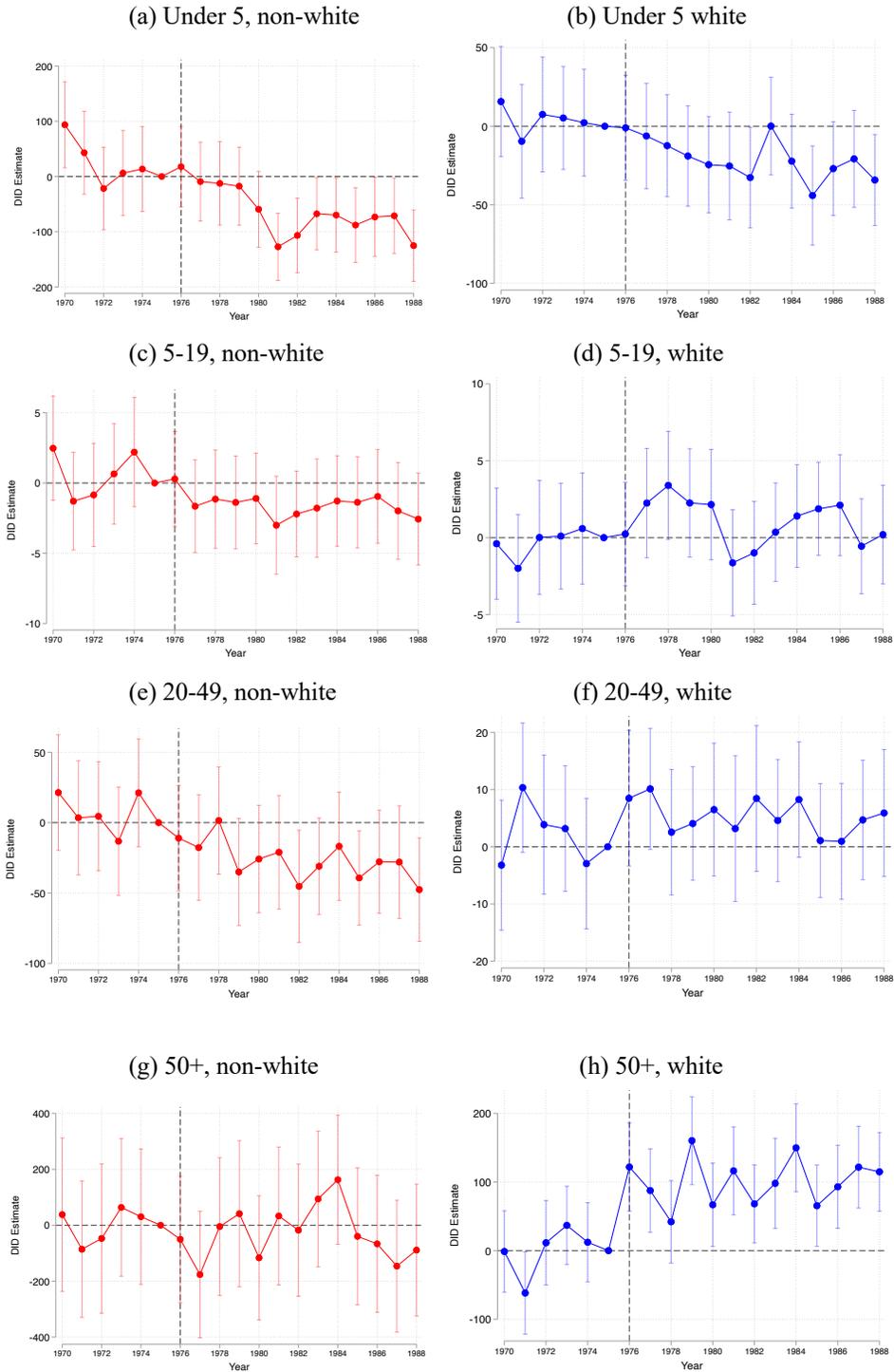

Notes: Estimates from Equation (2) of the main text using two-way fixed effects (TWFE) models. See Figure 2 notes for further details.

**Table A1**. ICD-8 and -9 codes for causes of death

| Cause of death | ICD-8 | ICD-9 |
|---|---|---|
| Cardiovascular diseases | 390-398 | 390-398 |
|  | 400-404 | 401-404 |
|  | 410-413 | 410-414 |
|  | 420-429 | 415-429 |
|  | 440 | 440 |
|  | 441-448 | 441-448 |
| Cerebrovascular disease (stroke) | 430-438 | 430-438 |
| Cancer | 150-163 | 150-165 |
|  | 174 | 174-175 |
|  | 180-189 | 179-189 |
|  | 204-207 | 204-208 |
| Infectious disease | 0-136 | 1-18 |
|  |  | 20-88 |
|  |  | 90-139 |
| External causes (e.g., accidents, injuries, suicide) | 800-999 | 800-999 |

Notes: We use publicly available NCHS multiple cause of death files from 1970-1988 and county intercensal population estimates from the U.S. Census Bureau to calculated race stratified (white, non-white) county-level all-cause mortality rates per 100,000 population. We used the 1970 population stratified by 5-year age bins to standardize all mortality rates. We calculated cause-specific age-adjusted mortality rates for each race group per 100,000 population for cardiovascular disease, cerebrovascular disease, cancer, infectious disease, and external causes using the ICD-8 and -9 codes in the table above.

**Table A2**. SDID time weights

|  | Non-white | | | | White | | | |
|---|---|---|---|---|---|---|---|---|
|  | Under-5 | 5-19y | 20-49y | 50+ | Under-5 | 5-19y | 20-49y | 50+ |
| 1970 | .1229488 | .1567752 | .1290679 | .1405993 | .1695963 | .1557951 | .1646454 | .1693898 |
| 1971 | .1436308 | .1100265 | .1291887 | .1413844 | .1325056 | .1551328 | .1742084 | .1188578 |
| 1972 | .1313393 | .2165676 | .1672041 | .1291859 | .1073533 | .1496477 | .1081166 | .0996935 |
| 1973 | .2029925 | .1927311 | .1777515 | .170504 | .1729463 | .1736873 | .1822966 | .1680776 |
| 1974 | .196034 | .1478055 | .1562942 | .190236 | .1972152 | .1759313 | .174044 | .2002697 |
| 1975 | .2030546 | .1760942 | .2404936 | .2280904 | .2203832 | .1898059 | .1966889 | .2437116 |

Notes: Estimates of SDID time weights from Equation (1).

## Table A3. TWFE estimates and population weights

| | Non-white | | | | White | | | |
|---|---|---|---|---|---|---|---|---|
| | (1) | (2) | (3) | (4) | (5) | (6) | (7) | (8) |
| | SDID | TWFE | TWFE with pop. weights | TWFE with pop. weights, accounting for heterogeneity | SDID | TWFE | TWFE with pop. weights | TWFE with pop. weights, accounting for heterogeneity |
| **Under 5** | -80.82*** | -81.28*** | -61.63*** | -98.64*** | -21.11*** | -23.58*** | -32.92*** | -31.5*** |
| | (17.91) | (15.96) | (15.96) | (24.9) | (5.80) | (6.43) | (8.26) | (6.22) |
| | | | *608.3* | | | | *326.7* | |
| **5-19y** | -2.15*** | -2.06*** | -0.12 | -2.87* | 1.038 | 1.328** | -0.0692 | 0.80 |
| | (0.65) | (0.66) | (1.52) | (1.52) | (0.71) | (0.67) | (0.43) | (0.69) |
| | | | *36.7* | | | | *21.4* | |
| **20-49y** | -30.53*** | -32.26*** | -16.38* | -23.0** | 3.69 | 3.56 | 1.38 | 2.38 |
| | (8.05) | (7.90) | (9.41) | (11.87) | (2.56) | (2.32) | (1.96) | (2.13) |
| | | | *368.3* | | | | *151.4* | |
| **50 and above** | -18.42 | -33.15 | 16.00 | -11.22 | 98.47*** | 100.4*** | 47.69*** | 90.45*** |
| | (58.99) | (52.91) | (27.19) | (46.39) | (16.17) | (15.55) | (11.50) | (12.81) |
| | | | *3068.1* | | | | *2904.7* | |

Notes: Each cell represents a separate regression. Columns (1) and (4) report the core SDID estimates of Equation (1). Columns (2) and (6) present estimates of the same equation using two-way fixed effects (TWFE) models. Columns (3) and (7) estimate the same TWFE models but weight regressions by race-age group population. Columns (4) and (8) estimate Equation (1) using weighted TWFE models and additionally including interactions between all RHS variables and binary indicator for 1970 county population in the top decile of all sample counties (~>125,000). The coefficients presented are on the main effect of $Post_t \times CoveredVRA_i$. We use this model to assess whether the treatment effect heterogeneity implied by the difference between weighted and unweighted TWFE models can be accounted for by differences in treatment effects by area-level population. Consistenet with this possibility, we find that the estimates in Cols (4) and (8) are generally much closer to the main SDID estimates and unweighted TWFE estimates than they are to the weighted TWFE estimates. See main text for further details.
\*\*\* - p<0.01, \*\* - p<0.05, \* - p<0.1.

**Table A4**. Triple difference estimates

|  | DIDID Estimates |
|---|---|
| **Under 5** | -56.8*** |
|  | (16.84) |
| **5-19y** | -3.29*** |
|  | (0.94) |
| **20-49y** | -36.1*** |
|  | (8.20) |
| **50 and above** | -119.8** |
|  | (52.23) |

Notes: Each row represents a separate triple differences regression, estimated using two-way fixed effects. The model is identical to Equation (1) except, for each age group, we pool observations for both white and non-white persons and interact a binary indicator denoting observations for non-white individuals against all RHS variables. In addition, to assess potential county-year confounds that affect both racial groups, we include county-year fixed effects in these models. The coefficients presented are for the triple difference term ($Post_t \times CoveredVRA_i \times Non-White$).
*** - p<0.01, ** - p<0.05, * - p<0.1.

**Table A5**. Cause-specific estimates by race and gender

| | | Non-white | | | | | White | | | |
|---|---|---|---|---|---|---|---|---|---|---|
| | (1) | (2) | (3) | (4) | (5) | (6) | (7) | (8) | (9) | (10) |
| | Cardio. | Cerebrovasc. | Cancer | Infectious | Accidental | Cardio. | Cerebrovasc. | Cancer | Infectious | Accidental |
| **Men** | | | | | | | | | | |
| 20-49y | 0.81 | -0.23 | -1.96*** | -0.05 | -11.6** | 3.88* | -0.07 | -1.60 | 0.02 | 4.37 |
| | (2.54) | (0.27) | (0.56) | (0.05) | (4.50) | (2.06) | (0.41) | (1.12) | (0.14) | (2.97) |
| | *104.3* | *9.95* | *37* | *4.1* | *114.9* | *71.3* | *7.65* | *26.6* | *1.9* | *76.8* |
| 50 and above | 18.93 | -36.25*** | -10.20 | 1.67 | -3.80 | 70.9*** | 22.36*** | -29.1*** | 2.35*** | -0.18 |
| | (34.2) | (11.53) | (17.5) | (1.18) | (4.70) | (14.70) | (5.91) | (6.43) | (0.75) | (3.12) |
| | *1527.5* | *403.7* | *683.3* | *44.1* | *134.5* | *1775.3* | *354.4* | *583.1* | *21* | *105* |
| **Women** | | | | | | | | | | |
| 20-49y | -1.58 | -0.28 | -1.57* | -- | -0.19 | 0.78 | 0.25 | -1.41 | -0.02 | 0.30 |
| | (1.15) | (0.20) | (0.88) | -- | (0.79) | (0.90) | (0.38) | (1.09) | (0.05) | (0.98) |
| | *51.5* | *8.42* | *42.5* | *--* | *28.3* | *19.3* | *7.82* | *36.7* | *1.5* | *21.1* |
| 50 and above | -32.70 | -49.69*** | -10.51 | 0.53 | -1.35 | 75.30*** | 10.82* | -13.04** | -0.37 | -1.56 |
| | (31.83) | (10.78) | (7.91) | (0.62) | (1.34) | (14.79) | (5.64) | (5.47) | (0.70) | (1.55) |
| | *1111.1* | *384.9* | *383.5* | *21.4* | *44.1* | *1156.1* | *361.7* | *384.5* | *12.9* | *59.9* |

Notes: Each column-panel presents SDID estimates of the average treatment effects of VRA coverage on mortality from a specific cause for a given race-age-gender group. See Table 4 notes for further details. Estimates for infectious disease mortality for non-white women ages 20-49 could not be computed given the low frequency of deaths recorded from this cause in this group.
*** - p<0.01, ** - p<0.05, * - p<0.1.

**Table A6**. VRA coverage and household income for selected sub-groups

|  | Non-white | | White | |
|---|---|---|---|---|
|  | (1) | (2) | (3) | (4) |
|  | Above 50 | <HS | Above 50 | <HS |
| **Ln(Family Income)** | 0.16 | 0.23** | 0.12** | 0.08*** |
|  | (0.099) | (0.11) | (0.033) | (0.018) |
| **Family Income** | 2,730*** | 4,610*** | 2,280 | 1,880*** |
|  | (1,010) | (1,420) | (2,520) | (558) |

Notes: Each cell represents a separate regression using CPS-ASEC data. Models are identical to those estimated in Table 6 of the main text except here we focus on specific sub-populations, individuals over 50 or those reporting completing less than a high school education. See Table 6 notes for further details.
*** - $p<0.01$, ** - $p<0.05$, * - $p<0.1$.